\documentclass{article}

\usepackage{arxiv}

\usepackage[utf8]{inputenc} % allow utf-8 input
\usepackage[T1]{fontenc}    % use 8-bit T1 fonts
\usepackage{hyperref}       % hyperlinks
\usepackage{url}            % simple URL typesetting
\usepackage{booktabs}       % professional-quality tables
\usepackage{amsfonts}       % blackboard math symbols
\usepackage{nicefrac}       % compact symbols for 1/2, etc.
\usepackage{microtype}      % microtypography
\usepackage{lipsum}		% Can be removed after putting your text content
\usepackage{graphicx}

\usepackage{amsmath}
\usepackage{algorithm}
\usepackage{algorithmicx}
\usepackage{algpseudocode}
\usepackage{subfigure}

\title{Adversarial Robustness of Deep Convolutional Candlestick Learner}

\author{Jun-Hao Chen\\
	Department of Computer Science \& Information Engineering\\
	National Taiwan University\\
	Taipei 10617, Taiwan\\
	\And
	Samuel Yen-Chi Chen\\
	Computational Science Initiative\\
	Brookhaven National Laboratory\\
	Upton, NY 11973, USA\\
	\And
	Yun-Cheng Tsai\\
	School of Big Data Management\\
	Soochow University\\
	Taipei 11102, Taiwan\\
	\texttt{pecutsai@gm.scu.edu.tw} \\
	\And
	Chih-Shiang Shur\\
	Department of Business Administration\\
	National Sun Yat-sen University\\
	Taipei 10617, Taiwan
}

\begin{document}
\maketitle

\begin{abstract}
Deep learning (DL) has been applied extensively in a wide range of fields. However, it has been shown that DL models are susceptible to a certain kinds of perturbations called \emph{adversarial attacks}. To fully unlock the power of DL in critical fields such as financial trading, it is necessary to address such issues. In this paper, we present a method of constructing perturbed examples and use these examples to boost the robustness of the model. Our algorithm increases the stability of DL models for candlestick classification with respect to perturbations in the input
data.
\end{abstract}

% keywords can be removed

\section{Introduction}
\label{sec:introduction}
Deep learning (DL) has achieved incredible success in recent years. Fields like image recognition and natural language processing dominated by DL and relevant techniques have been adopted in industry extensively~\cite{redmon2018yolov3, devlin2018bert}. Despite the powerful capability, deep learning still has some disadvantages, such as instability and are easy to fool. Many studies have shown that deep learning models can change misl easily by crafted examples like performing slight perturbations in data that human eyes cannot even detect~\cite{szegedy2013intriguing, nguyen2015deep}. This vulnerability of the artificial intelligence system would make a person or cooperate in suffering from significant loss.

When applying deep learning in the financial industry, the vulnerability of the artificial intelligence system is more critical since any misleading could result in wrong decisions that probably cause millions or billions loss. Therefore, it is urgent to improve the stability and noise tolerance to extinguish suspicion to deep learning and artificial intelligence system.

Adversarial training is one of the methods to improve the stability of deep learning models. Several studies have demonstrated that using adversarial examples during the training of deep learning models could improve both classification accuracy and stability on test data~\cite{szegedy2013intriguing,  goodfellow2014explaining, shaham2018understanding}. In general, adversarial samples acquired from the output with perturbations that lead to misclassification. However, financial data is more sensitive than regular image data. Conventional perturbations will not make an image look different but will cause a dramatic change in financial data, which humans can easily find.

Regarding the issue mentioned above of adversarial training, we propose a novel Modified Local Search Attack Sampling model based on the previous work on explainable candlestick learners to develop an adversarial training method for financial candlestick data~\cite{chen2020explainable}. The model consists of three steps: 
\begin{enumerate}
\item \emph{Train a GAF-CNN model:}\\
    The GAF-CNN model is used to classify financial candlestick data. This method includes two processes: (a) encode time-series data to the GAF matrix and (b) train CNN model with GAF matrix~\cite{tsai2019encoding}.
\item \emph{Generate Adversarial Examples:}\\
    Applying the Modified Local Search Attack Sampling method to the trained model and generating examples with perturbations will not lead to misclassification.
\item \emph{Adversarial training:}\\
    We are merging adversarial examples with original clean examples to be a new dataset to train the new GAF-CNN model. 
\end{enumerate}
The central concept in our proposed method is that we adopt perturbed data, which is not misclassified by trained GAF-CNN to do adversarial training. The generated adversarial examples based on original clean examples with small perturbations, so they only have a slight difference and remain the characteristics of the original time series.

To evaluate our model's performance, we respectively train GAF-CNN model on two data: (1) clean and (2) merged examples, and compare the accuracy and stability. Merged examples compose of clean and perturbed examples. Besides, we performed statistical analysis to verify our conjecture. Since there is an intersection between clean and merged examples, we suppose the two populations of the output are dependent, and the dependent paired $t$-test adopted to analyze the result. We expect the model trained by merged examples has significantly better accuracy and stability than the one trained only with clean examples.

\section{Background}
Detecting the entry and exit points of the trading market is an important issue in finance. Crucial information for decision making often hides in the depths of the fluctuated market prices. To obtain this information more effectively, Munehisa Homma proposed a visually intuitive tool, \emph{Candlestick Chart}, to better observe the market situation behind the rice market in $18$th century~\cite{nison2001japanese}. A candlestick chart contains multiple bars, each bar displays open, high, low, and close price (OHLC) for a specific period. The period of a bar can be arbitrarily customized, usually depending on the length of the transaction. The OHLC data is illustrated in detail as follows:

\begin{enumerate}
    \item Open: the first price of a specific period.
    \item High: the highest price of a specific period.
    \item Low: the lowest price of a specific period.
    \item Close: the last price of a specific period.
\end{enumerate}

\begin{figure}
\centering
\includegraphics[scale=0.45]{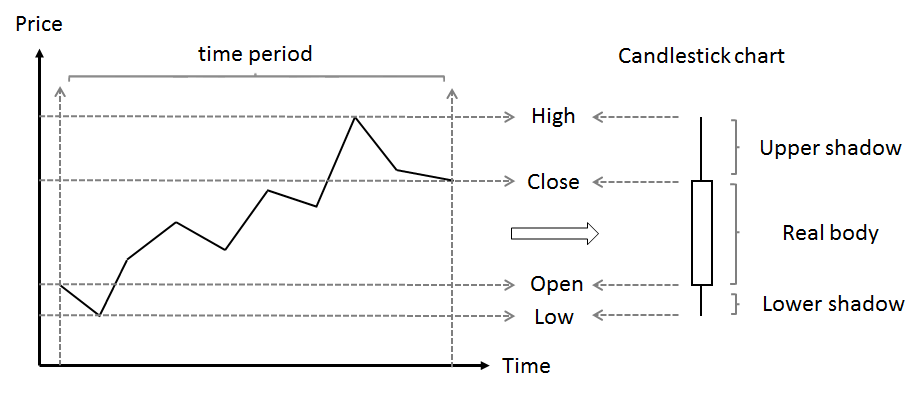}
\caption{Candlesticks display all the market needed information, such as open, high, low, and close prices.}
\label{candlestick_intro}
\end{figure}

Comparing to pure OHLC data, \emph{Real Body} and \emph{Shadows} are more often reference by traders since they can depict the characteristics of the bar more visually. The real body defined as the area between open and close prices. It will be filled by black or red if the open price is higher than the close price. Inversely, it will be filled by white or green. Shadows represent the lowest and highest prices during the specific period. All mentioned terminologies often used in traders’ discussions on the practical. Figure~\ref{candlestick_intro} illustrates the entire components of the candlestick chart in detail.

\begin{figure}
\centering
\includegraphics[width=0.45\textwidth]{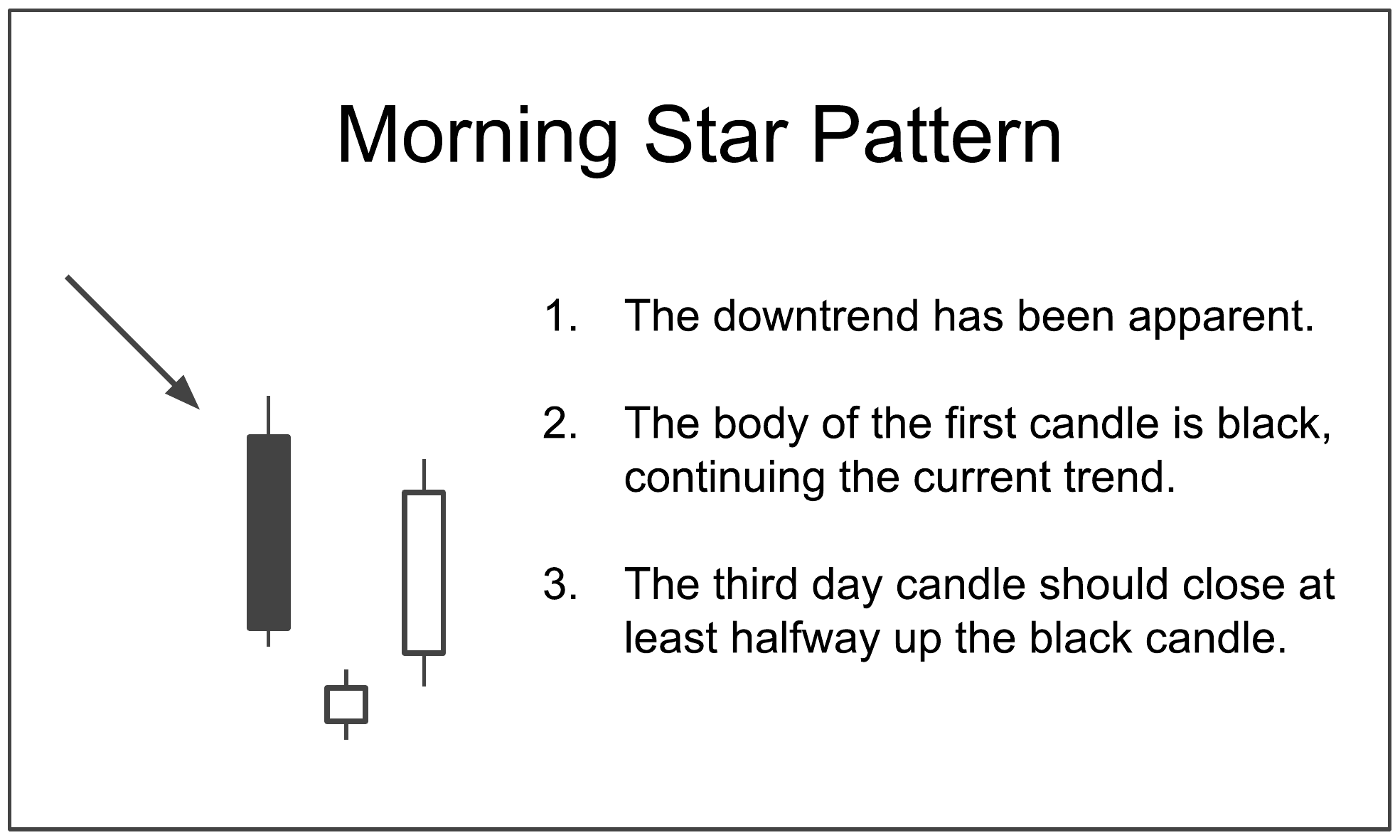}
\caption{{\bfseries Illustration of Morning Star Pattern.} The left-hand side shows the appearance of the Morning Star pattern. The right-hand side shows the critical rules of the Morning Star pattern.}
\label{morning_intro}
\end{figure}

\begin{figure}
\centering
\includegraphics[width=0.45\textwidth]{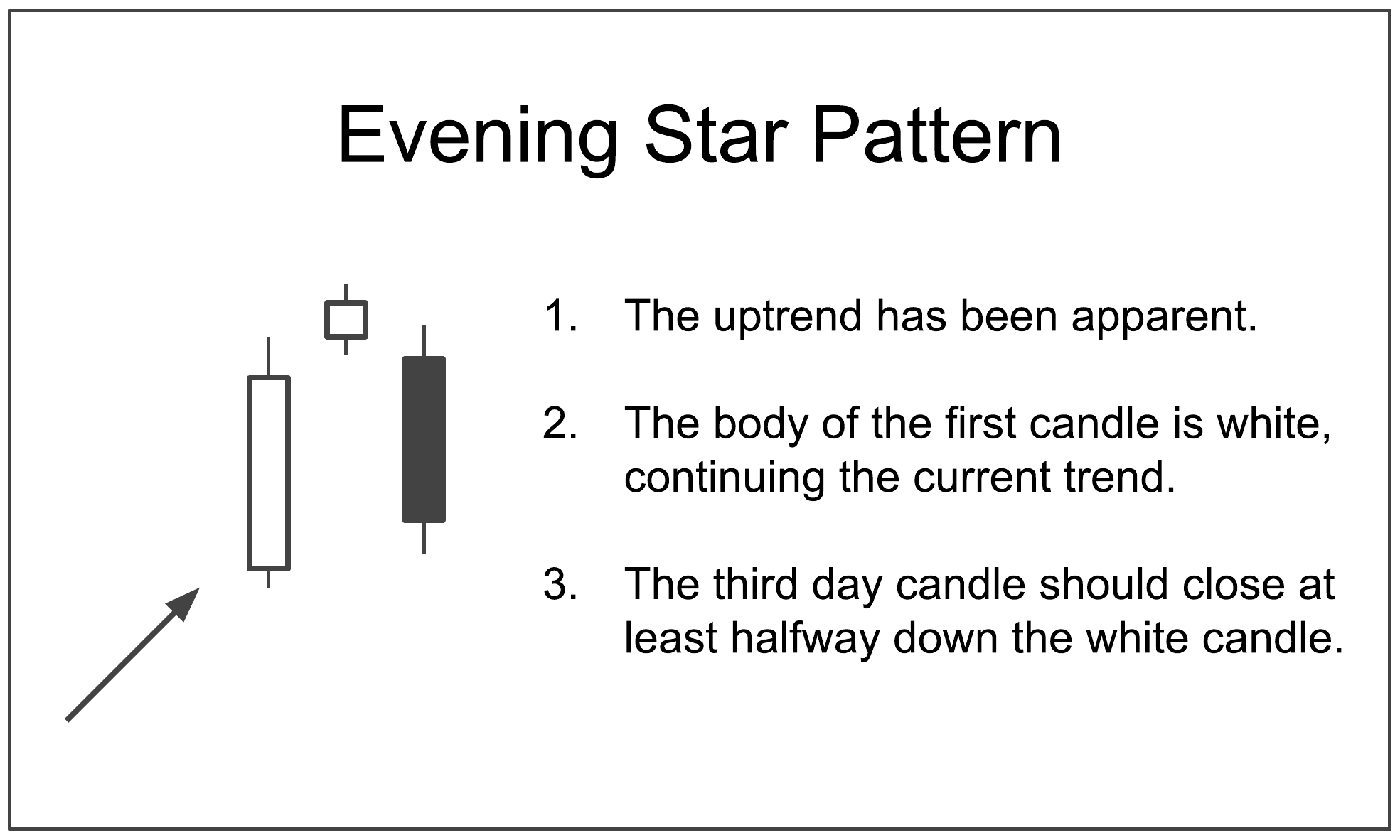}
\caption{{\bfseries Illustration of Evening Star Pattern.} The left-hand side shows the appearance of the Evening Star pattern. The right-hand side shows the critical rules of the Evening Star pattern.}
\label{evening_intro}
\end{figure}

In candlestick theory, the candlestick chart makes sense only if it has a particular combination, generally called \emph{Candlestick Pattern}. The candlestick pattern can help traders to capture the patterns of supply and demand behind the financial market~\cite{lee2006pattern}. \emph{The Major Candlesticks Signals} refers to several basic patterns of candlestick~\cite{MajorSignals}, and it refers to a pattern composed of two parts: 
\begin{enumerate}
    \item \emph{Trend:}\\
    The trend indicates a dynamic situation of price in a period that includes uptrend and downtrend.
    \item \emph{Three bars pattern:}\\
    Three bars pattern holds plenty of detail of characteristics for every pattern.
\end{enumerate}
Besides, referring to \emph{The Major Candlesticks Signals}, eight of these patterns are adopted in this study, including: 
\begin{enumerate}
    \item Morning Star is a bullish candlestick pattern. It could recognize by a downtrend followed by three bars pattern: a long black bar, a shorter black or white bar with a long lower shadow, and a long white bar. The middle bar of the morning star captures market indecision feature where the bears begin to give way to bulls. The third bar confirms the reversal and begins a new uptrend. The longer the third bar, the stronger the uptrend will be. Figure~\ref{morning_intro} illustrates the morning star pattern based on the description.
    \item Evening Star is a bearish candlestick pattern. It could recognize by an uptrend followed by three bars pattern: a long white bar, a short-bodied bar, and a black bar. The middle bar captures market indecision feature where the bulls begin to give way to bears. The third bar also confirms the reversal and begins a new downtrend. It will be more obvious with a long black bar. Figure~\ref{evening_intro} shows the evening star based on the description.
    \item Bullish Engulfing could recognize by a downtrend followed by three bars pattern: a short black bar, a long white bar engulfs the previous bar, and a white bar confirms the reversal.
    \item Bearish Engulfing could recognize by an uptrend followed by three bars pattern: a short white bar, a long black bar engulfs the previous bar, and a black bar confirms the reversal.
    \item Shooting Star could recognize by an uptrend followed by three bars pattern: a long white bar, a short bar with long upper shadow, and a black bar confirms the reversal.
    \item Inverted Hammer could recognize by a downtrend followed by three bars pattern: a long black bar, a short bar with a long lower shadow, and a white bar confirms the reversal.
    \item Bullish Harami could recognize by a downtrend followed by three bars pattern: a long black bar, a short white bar is engulfed by the previous bar, and a white bar confirms the reversal.
    \item Bearish Harami could recognize by an uptrend followed by three bars pattern: a long white bar, a short black bar is engulfed by the previous bar, and a black bar confirms the reversal.
\end{enumerate}

\section{Literature review}
With the ability of high expressiveness, deep learning models get extremely high achievement in recent years. However, behind the high accuracy, deep learning models still suffer from the risk of instability. In $2013$, Szegedy et al. first proposed the concept of \emph{adversarial examples}. They found out that they could easily mislead the well-trained model by applying imperceptible non-random perturbations, where the perturbations were generated by limited memory Broyden–Fletcher–Goldfarb–Shanno algorithm (L-BFGS)~\cite{szegedy2013intriguing}. Besides, Nguyen et al. also demonstrated that it is easy to produce images completely unrecognized by humans, but the state-of-the-art deep learning models recognize it with $99.99$\% confidence~\cite{nguyen2015deep}.

Regarding the problem mentioned above, Szegedy et al. first came up with the concept of the \emph{adversarial training}. They trained the neural network on a mixture of adversarial and clean examples, and the model ended up with higher stability but also higher error~\cite{szegedy2013intriguing}. In 2014, Goodfellow et al. implemented adversarial training more efficiently. They proposed a computationally efficient Fast Gradient Sign Method (FGSM) to generate the adversarial examples. With FGSM, they increase both accuracy and stability on MNIST dataset~\cite{goodfellow2014explaining}. Besides, Shaham et al. also proposed an adversarial training framework by using the minimization-maximization procedure. Both accuracy and stability improved on a more complex CIFAR-$10$ test dataset~\cite{shaham2018understanding}.

We apply the adversarial training more reasonably in financial candlestick data; this work is based on the contribution of these studies and proposes a new training framework for the GAF-CNN model.

\section{Methods}
\subsection{GAF-CNN}
The GAF-CNN is a two-step model that could effectively perform classification tasks on time series data. First, encode the time series data into a two-dimensional matrix with Gramian Angular Field (GAF) method. Second, use this two-dimensional matrix as input to train the Convolutional Neural Network (CNN) model with architectures based on the complexity of the specific task~\cite{wang2015imaging}. Through the GAF encoding method, the two-dimensional matrix we get is image-like, which allows the CNN model to \emph{visually} recognize, classify, and learn the underlying structures and patterns.   

The GAF is a novel time-series encoding method proposed by Wang and Oates~\cite{wang2015imaging}. The GAF method represents time-series data in the polar coordinate system and converts angles into the asymmetric two-dimensional matrix. GAF contains two different forms, summation and difference version, and the summation version adopted in this study. Each element of the GAF matrix is the cosine of the summation of angles. The summation between each point represents the temporal correlation within different time intervals.

First, normalize the time series data $X$ into values in $[0, 1]$. The following equation shows the entire normalization process. Notation $\widetilde{x}_{i}$ represents the normalized data.

\begin{align}
\widetilde{x}_{i}&=\frac{x_i-\min(X)}{\max(X)-\min(X)}
\end{align}

Second, represent the normalized time series data in the polar coordinate system, calculating the angles of normalized time series data $\widetilde{x}$ with the following two equations.

\begin{align}
\phi&=\arccos(\widetilde{x}_i), -1\leq \widetilde{x}_i \leq 1, \widetilde{x}_i \in \widetilde{X}\\
r&=\frac{t_i}{N}, t_i\in\mathbb{N}
\end{align}

Finally, sum the angles with cosine function to make the GAF matrix by the following equation:

\begin{align}
\textup{GAF}=\cos(\phi_i + \phi_j) = \widetilde{X}^T \cdot \widetilde{X} - \sqrt{I-\widetilde{X}^2}^T\cdot \sqrt{I-\widetilde{X}^2}
\end{align}

There are two essential properties of GAF. First, the mapping from the normalized time series data to GAF is bijective when $\phi \in [0,\pi]$. Therefore, the GAF matrix can inversely transform to time series data through its diagonal elements. Second, unlike Cartesian coordinates, the polar coordinates preserve absolute temporal relations. 

CNN model can effectively extract the spatial information by its local connectivity. This feature leads to high accuracy and is suited for tasks that rely on recognition and classification with a flexible architecture. Through GAF based encoding, which provides image-like property, we can utilize the power of CNN to perform the further study on time series data.  

\subsection{Modified Local Search Attack Sampling}
\begin{figure}
\begin{center}
\includegraphics[scale=0.4]{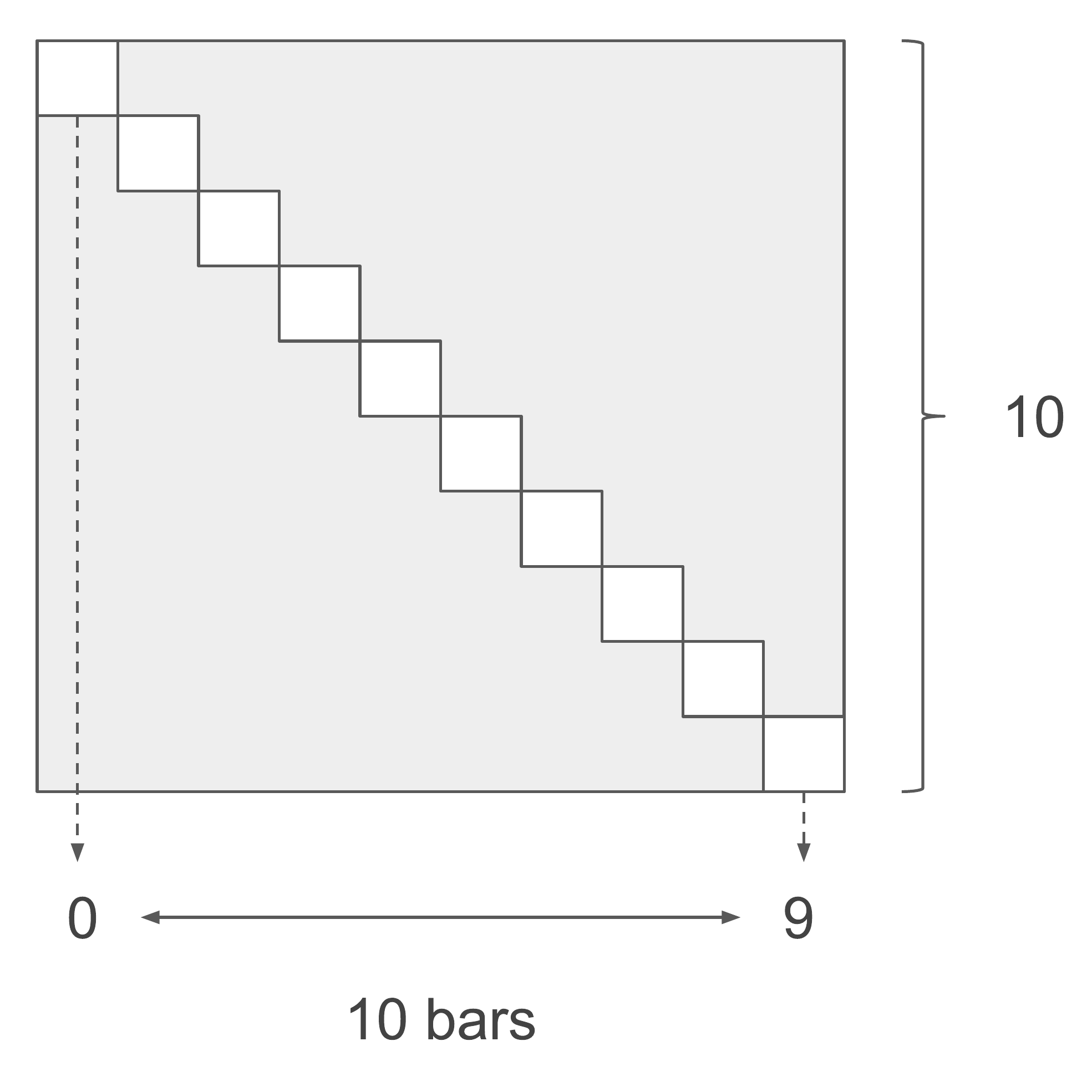}
\caption{The GAF matrix for the Modified Local Search Attack Sampling method. In this work, we attack the diagonal elements as these $10$ locations.}
\label{fig:attack_region}
\end{center}
\end{figure}

For realizing adversarial training on the GAF-CNN framework more reasonably, algorithms should modify to suit the characteristic of financial candlestick data. Modified Local Search Attack Sampling model designed to attack the data under the GAF-CNN framework, its operation will focus on the diagonal elements of the matrix, and the attack region is controllable~\cite{chen2020explainable}. Figure~\ref{fig:attack_region} shows the exact locations we attack. Through the GAF matrix's invertibility, it could easily transform back to candlestick time-series data that humans can recognize. We expect that adversarial examples of candlestick can keep the same class of pattern after perturbation while improving the stability and accuracy of the model.

In financial candlestick data, any subtle difference in the data can lead to different classification outcomes. For example, if perturbation turns any bar from white to black in a candlestick pattern, it will destroy the original pattern's properties and lead to the perceptible difference from the original data to the human trader. Any difference, even a slight perturbation, would make the generated adversarial examples violate the rules defined by the human trader and loss what it should have to make it a pattern.

For controlling the perturbations to prevent the misclassification, we make two modifications to the original Modified Local Search Attack model. First, we downscale the random perturbation scale from uniform distribution to the interval $[0.99, 1.01]$, which further limits the range of perturbations with the original reset mechanism. Second, instead of collecting adversarial examples from successful attacks, we collect them from failed attacks. Adversarial examples from failed attacks have higher chances to follow the rules that we defined. On the other hand, adversarial examples of candlestick patterns cannot contradict the intuition of traders.

The details of the algorithm are in Algorithm~\ref{alg}. First, perturb the diagonal elements of the GAF matrix in the range of uniform distribution $[0.99, 1.01]$, and reset it every three episodes. The purpose of a reset here is to limit the extent of perturbation. Based on the perturbed diagonal elements of the GAF matrix, we then calculate the corresponding values of non-diagonal elements and output the perturbed GAF matrix. If the CNN model does not misclassify the perturbed GAF, we collect it as an adversarial example. Otherwise, we repeat the process until reaching $R$ times. Furthermore, the entire collected adversarial examples include all intermediate perturbed data, for example, when a parameter of reset is $3$, and an adversarial example perturbed three times, all examples include one time, two times and three times perturbed adversarial examples will be collected. 

\begin{algorithm}[tb]
\begin{algorithmic}

\State Load a single GAF two-dimensional array $A$
\State Set $T = \text{length of the time-series}$ 
\State Keep a copy of $A$ in memory $D$
\State Initialize the counter $t = 0$
\For{episode $=1,2,\ldots,R$}

\If{$t = 3$}
\State Reinitialize the $A$ to the original value from memory $D$
\State Reset the counter $t = 0$
\EndIf
\For {$l = 1,2,\ldots,T$}
    \State Sampling a random perturbation scale $r_l$ from uniform distribution $[0.99,1.01]$
    \State Calculate the perturbed result $=r_l \times A[l,l]$
    \State Set $A[l,l] = r_l \times A[l,l]$ 
\EndFor

\State $t = t + 1$
\State Recalculate the time series from perturbed $A$ and then encode into a new GAF matrix $A'$
\If{$A'$ \text{is not adversarial}}
    \State Collect $A'$ as generated data
\EndIf
\EndFor
\end{algorithmic}
\caption{Modified Local Search Attack Sampling}
\label{alg}
\end{algorithm}

\section{Experiments}
\subsection{Clean examples creation}
We use foreign exchange EUR/USD $1$-minute OHLC data, from January 1, 2010, to January 1, 2018, to label the financial candlestick patterns. We reference The Major Candlestick Signals to design the rules of candlestick patterns and choose eight patterns to label~\cite{MajorSignals}. The rules in each class composed of a trend and a three bars pattern. Finally, the entire clean examples contain eight patterns, and each label includes $1500$ labeled data.

\subsection{Adversarial examples creation}
\begin{figure}
\begin{center}
\includegraphics[width=1.0\linewidth ]{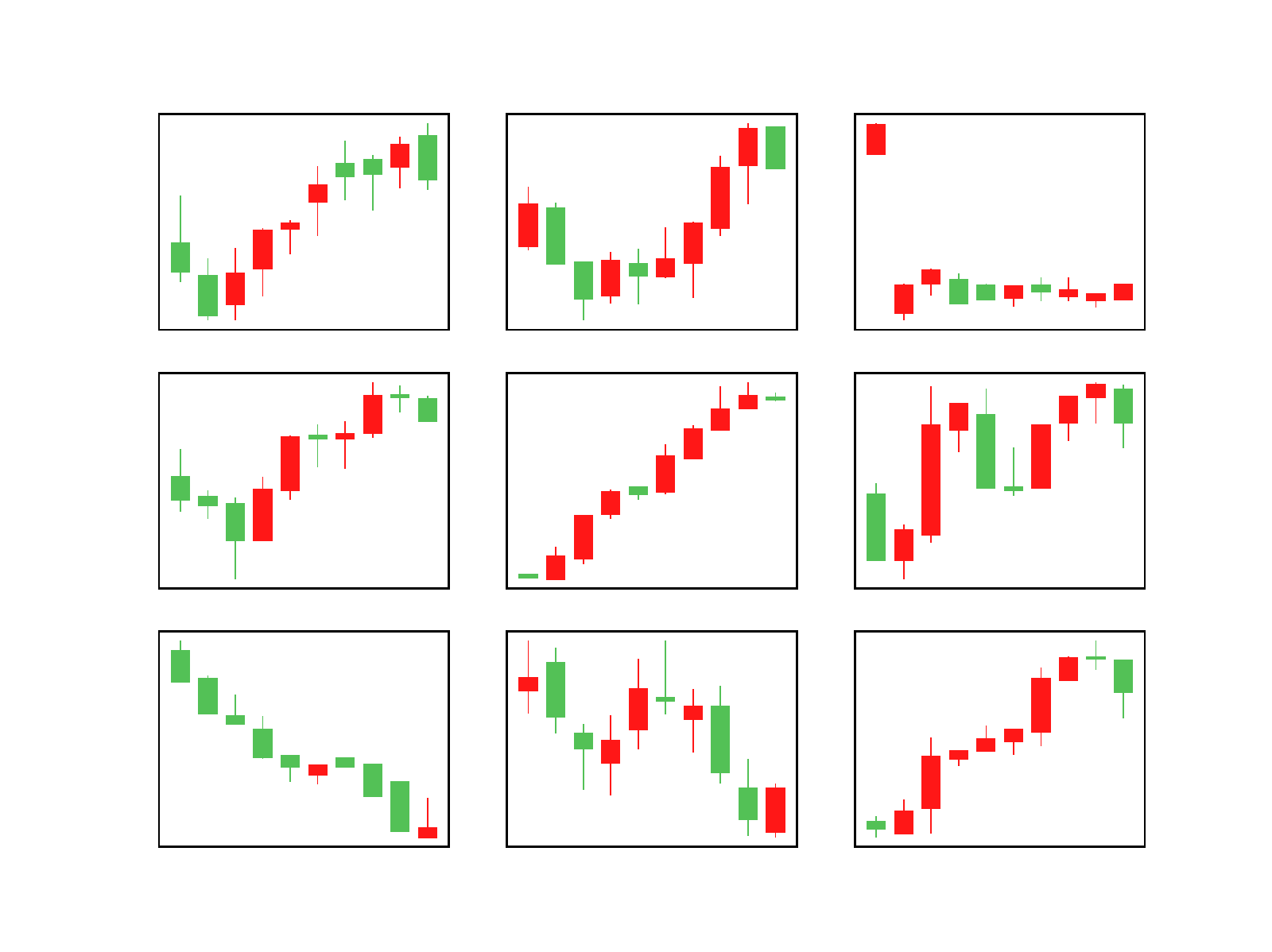}
\end{center}
\caption{The example data generated by the Modified Local Search Attack Sampling model.}
\label{fig:adv_examples}
\end{figure}

We transform clean examples to matrix form through the GAF encode method, then slightly perturb diagonal elements of the matrix through the Modified Local Search Attack Sampling method. Since the algorithm collects all accumulated perturbed data, collected adversarial examples are far more than clean examples. We randomly sample adversarial examples so that the merged dataset contains $50$\% clean and $50$\% adversarial examples.  Figure~\ref{fig:adv_examples} shows nine random candlestick patterns from adversarial examples.

\subsection{Adversarial training}
To evaluate whether adversarial training can effectively improve the accuracy and stability of the model, we respectively train GAF-CNN $100$ times with clean and merged examples under the same parameters. We perform a dependent paired $t$-test on the accuracy to infer that the accuracy of merged examples is significantly different to clean examples. Moreover, we respectively attack models trained by clean and merged examples then observe the successful attack ratio to evaluate the difference of stability.

\section{Results}
\subsection{Accuracy testing}
After $100$ times training with clean and merged examples, we compare the accuracy between each example, where $100$ accuracies each has, by statistical analysis. Table~\ref{tab:descriptive_statistics} shows the descriptive statistics of the model accuracies trained by both datasets. The mean accuracy of models trained with merged examples is $90.85$\%, higher than the ones solely trained with clean examples. The standard deviation of model accuracies trained with merged examples is $0.3687$, which is also lower than the ones trained by clean examples.

% Mean/std summary
\begin{table}[]
\centering
\begin{tabular}{cccc} \toprule
Model & Mean of Accuracies (\%) & Std of Accuracies (\%) \\ \midrule
clean examples & 90.72 & 0.3969 \\
merged examples & 90.85 & 0.3687 \\ \bottomrule
\end{tabular}
\caption{The descriptive statistics between 100 times trained by clean and merged examples.}
\label{tab:descriptive_statistics}
\end{table}

To perform the statistical inference, the dependent paired $t$-test adopted to analyze the training results. In the null hypothesis $H_0$, we assume that the accuracies have no significant difference. According to the results in Table~\ref{tab:ttest_table}, the two-tailed dependent paired $t$-statistic is $-2.5294$ and the $p$-value is $0.0130$. Since the $p$-value is less than $0.025$, this means that the accuracies of models trained by clean and merged data are significantly different. Regarding the fact that merged examples have a higher mean of accuracies, we can state that adversarial training via our Modified Local Search Attack Sampling model significantly improves the accuracy of the model.

% Paired T-value table
\begin{table}
\small
\centering
\def\arraystretch{1.3}
\begin{tabular}{ccclll} \toprule
\multicolumn{6}{c}{Dependent Paired T Test} \\
$H_0$ & N & Mean & Std & T-value & P-value \\ \midrule
$\mu_{\textup{clean}}=\mu_{\textup{merge}}$ & 100 & -0.0013 & 0.0051 & -2.5294 & 0.0130 \\ \bottomrule
\end{tabular}
\caption{Two sided test at the $0.05$ significance level.
The null hypothesis stated here is that the accuracies have no significant difference between models trained by clean and merged examples.}
\label{tab:ttest_table}
\end{table} 

\subsection{Stability testing}
We select the highest accuracy model in each example to be the attack target, where the model accuracy trained with a clean example is $91.74$\%, and the model accuracy trained with merged example is $91.78$\%. Then compare their success rate of each label and all data. The low success rate means high stability. 

Table~\ref{tab:clean_success_rate_table} and~\ref{tab:merged_success_rate_table} respectively show the success rate of attack on models trained with clean examples and merged examples. The total average success rate of attack on a model trained with merged examples is $62.64$\%, which is lower than $70.88$\% from the attack on the model trained with clean examples. All labels except label $4$ have a lower success rate with merged examples. Although the success rate of label $4$ is a little higher after adversarial training, the others have obvious improvement, where the total average success rate is $8.24$\% lower.  

\begin{table}[]
\centering
\begin{tabular}{ccc} \toprule
Label & Success Rate & Percent (\%) \\
\midrule
1     & 505          & 33.67        \\
2     & 952          & 63.47        \\
3     & 1238         & 82.53        \\
4     & 1408         & 93.87        \\
5     & 715          & 47.67        \\
6     & 1047         & 69.80        \\
7     & 1265         & 84.33        \\
8     & 1375         & 91.67        \\
\midrule
Avg   &              & 70.88        \\
\bottomrule
\end{tabular}
\caption{The attack success rate of best model trained by clean examples.}
\label{tab:clean_success_rate_table}
\end{table}

\begin{table}[]
\centering
\begin{tabular}{ccc} \toprule
Label & Success Rate & Percent (\%) \\
\midrule
1     & 482          & 32.13        \\
2     & 918          & 61.20        \\
3     & 1163         & 77.53        \\
4     & 1420         & 94.67        \\
5     & 392          & 26.13        \\
6     & 894          & 59.60        \\
7     & 965          & 64.33        \\
8     & 1283         & 85.53        \\
\midrule
Avg   &              & 62.64        \\
\bottomrule
\end{tabular}
\caption{The attack success rate of best model trained by merged examples.}
\label{tab:merged_success_rate_table}
\end{table}

\section{Discussion}
\subsection{Perturbation with time-series property}
In this study, we perform the same perturbations on the GAF matrix's diagonal elements, which implies all $10$ bars in candlestick data. However, the time-series data ordered, where the important characteristic of features depends on the ordering~\cite{bagnall2017great}. Hence, we could give different degrees or scales of perturbations according to the order. For example, it is possible to perturb more recent data with a smaller scale of perturbations.  Considering the properties of time series data, it could bring a more cognitive perturbation process, and lead to a more consistent adversarial training.

\subsection{Perturbation with input data property}
Besides the property of time series data, the different regions of input data have their characteristics, and it is worth investigating these finer details. We can assign different degrees or scales of perturbations in each region. For example, the trend (front $7$ bars) can tolerate more perturbations than the last three bars, representing the patterns, because of the vast difference between uptrend and downtrend. In other words, we could do more perturbations on the trend region while only with slight perturbations on the three bars pattern, to help model finding a better decision boundary for the better classification result.

\subsection{Explainable learner}
The perturbations of the different regions can also be used in explainable artificial intelligence (XAI)~\cite{gunning2017explainable}. We can observe the adversarial training results, where adversarial examples are from perturbation of either the trend region or the three bars pattern. If adversarial examples from the three bars pattern have a better result, it can imply that the three bars pattern contributes more to improving accuracy and stability. Besides, we can further study the effects of perturbations on each bar of the three bars pattern. In summary, this explainable learner can help us identify the features which model rely on and investigate whether the trained classifiers follow the rules defined by human traders.

\section{Conclusion}
% We provide an open-source implementation and training data for the paper in the following URL: \url{https://github.com/pecu/FinancialVision}.

In summary, this paper proposes a novel Modified Local Search Attack Sampling model that generates financial candlestick adversarial examples, which can boost the model performance both in accuracy and stability. We use merged examples which compose of adversarial examples and clean examples to do the adversarial training. Based on the $100$ results from both examples, we demonstrate that adversarial training can significantly improve the accuracy and stability of the model.

Based on the potential of improving stability and interpretability, the model we propose is more trustworthy than black-box models. Professional traders can develop a financial trading strategy system with high accuracy and low risk with our model. This work could be a representative practice that facilitates the integration of machine learning to the financial industry. The implementation of this study and the data provided in the following URL:\url{https://github.com/pecu/FinancialVision}.

\section*{Acknowledgment}
Jun-Hao Chen and Yun-Cheng Tsai are supported in part by the Ministry of Science and Technology of Taiwan under grant 108-2218-E-002-050-. Samuel Yen-Chi Chen is supported in part by the U.S. Department of Energy, Office of Science, Office of High Energy Physics program under Award Number DE-SC-0012704 and the Brookhaven National Laboratory LDRD \#20-024. Yun-Cheng Tsai and Samuel Yen-Chi Chen conceived of the presented idea. Jun-Hao Chen and Chih-Shiang Shur developed the theory and performed the computations. All authors verified the analytical methods and discussed the results and contributed to the final manuscript. Thanks to Prof. Jane Yung-Jen Hsu for constructive discussion and great support.

\bibliographystyle{unsrt}
\bibliography{references}

\end{document}